## Main Manuscript for

# Spin sensitive transport in $RuCl_3$/Pt heterostructures: revealing field-transverse spin anisotropy in a spin liquid candidate

H. Idzuchi[1-3*], M. Kimata[4], S. Okamoto[5,6], P. Laurell[7], N. Mohanta[5], M. Cothrine[8], S. E. Nagler[9], D. Mandrus[8], A. Banerjee[2,6*] and Y. P. Chen[1,2,6,10,11*]

[1] *WPI Advanced Institute for Materials Research (AIMR) and Center for Science and Innovation in Spintronics (CSIS), Tohoku University, Sendai 980-8577, Japan.*

[2] *Department of Physics and Astronomy and Purdue Quantum Science and Engineering Institute, Purdue University, West Lafayette, Indiana 47907, USA.*

[3] *Department of Physics, The University of Tokyo, Bunkyo-ku, Tokyo 113-0033, Japan.*

[4] *Institute for Materials Research (IMR), Tohoku University, Sendai 980-8577, Japan.*

[5] *Materials Science and Technology Division, Oak Ridge National Laboratory, Oak Ridge, Tennessee, 37831 USA.*

[6] *The Quantum Science Center, Oak Ridge National Laboratory, Tennessee, 37831 USA*

[7] *Computational Sciences and Engineering Division, Oak Ridge National Laboratory, Oak Ridge, Tennessee 37831, USA.*

[8] *Department of Materials Science and Engineering, University of Tennessee, Knoxville, Tennessee 37996, USA.*

[9] *Neutron Scattering Division, Oak Ridge National Laboratory, Oak Ridge, Tennessee 37831, USA.*

[10] *School of Electrical and Computer Engineering and Birck Nanotechnology Center, Purdue University, West Lafayette, Indiana 47907, USA.*

[11] *Institute of Physics and Astronomy and Villum Centers for Dirac Materials and for Hybrid Quantum Materials and Devices, Aarhus University, 8000 Aarhus-C, Denmark.*

***Email:** idzuchi@g.ecc.u-tokyo.ac.jp (H. Idzuchi), arnabb@purdue.edu (A. Banerjee), yongchen@tohoku.ac.jp (Y. P. Chen)

**This PDF file includes:**

Main Text
Figures 1 to 5

**Current affiliation:** Department of Physics and Astronomy, University of Missouri, Columbia, Missouri 65211, USA (P. Laurel). Department of Physics, Indian Institute of Technology Roorkee, Roorkee 247667, India (N. Mohanta).

## Abstract

Alpha-phase (α-) $RuCl_3$ has emerged as a possible candidate for a quantum spin liquid (QSL) that promises exotic quasi-particles relevant for fault-tolerant quantum computation. Here, we report spin-sensitive transport measurements using a proximal spin Hall metal – platinum (Pt) – to probe magnetic moments in the insulator α-$RuCl_3$. We observe a spin Hall magnetoresistance (SMR) where both the transverse and longitudinal resistivities of α-$RuCl_3$/Pt exhibit oscillations as functions of the angle between the in-plane magnetic field and the current, arising from the interplay between the spin Hall effect in Pt and local magnetic moments in α-$RuCl_3$. The oscillations are observed from 1.5 T to 18 T, spanning the range of magnetic fields where zig-zag antiferromagnetic phase, the putative QSL phase, and the partially field-polarized states are reported in α-$RuCl_3$. The phase of the SMR oscillations suggest that the local moments, regardless of static or fluctuating, in α-$RuCl_3$ develop an anisotropy with an in-plane quantization axis that aligns largely transverse to the magnetic field for all fields above 1.5 T. Temperature dependance of the SMR revealed that the spin anisotropy has a similar energy scale to reported QSL signatures in α-$RuCl_3$. The coupling between the spin states within α-$RuCl_3$ and Pt demonstrated in our experiment opens a transport route to exploring exotic spin phases and device functionalities of QSL materials.

## Main Text

## Introduction

Spin liquid materials, where spins or magnetic moments do not order down to the lowest temperatures, can host exotic quasi-particles and excitations related to the spin degrees of freedom (1,2). Lately, the honeycomb-lattice antiferromagnet alpha-phase (α-) $RuCl_3$ has attracted significant attention (3-6) as a promising candidate for the "Kitaev quantum spin liquid (QSL)" that can host exotic topological gapless and gapped excitations (7). Alpha-phase $RuCl_3$ is a strongly spin-orbit-coupled Mott insulator with a relatively large charge gap of ~1eV, impeding observing effects from the charge degrees of freedom at low temperatures. The material features the anisotropic bond-dependent nearest-neighbor Kitaev coupling ($K_1$) as the leading order exchange coupling, resulting in strong bond frustration (3). The phase diagram in α-$RuCl_3$ has been experimentally characterized by several methods including thermodynamic measurements (such as heat capacity (4,8,9)), static magnetization (10), and neutron scattering (4,11) measurements. The spin degrees of freedom can support gapped or gapless excitations depending on the external magnetic field. For an ideal Kitaev system, it is suggested that when subject to a magnetic field in the appropriate direction, the zero-field gapless phase can turn into gapped phases hosting non-Abelian anyons, which are highly sought after as a means for topological quantum computing (7,12). Experimentally, nuclear magnetic resonance and neutron scattering measurements in α-$RuCl_3$ (13-15) under finite magnetic fields have been suggested to be consistent with a gapped QSL state. Recent experiments further reported a quantized thermal transport supporting the presence of a chiral Majorana Fermion edge mode expected in such a gapped topological QSL phase under a magnetic field (16-19). While the α-$RuCl_3$ local moments in zero and low fields order in a zigzag antiferromagnetic (AFM) state (4,8,11,20,21), the nature of the finite-field phase, where the AFM order is suppressed, including the QSL phase, remains inconclusive as it is challenging to probe the delicate spin properties when the ordered moments are small or vanishing. Theoretically, the full 3D Hamiltonian which incorporates all the in- and out-of-plane interactions, is also unresolved.

In this work, we exploit spin Hall effects (SHEs) (22) in the strongly spin-orbit-coupled metal platinum (Pt) to electrically probe the spin properties of α-$RuCl_3$ and the dependence on the magnetic field and temperature. In particular, our measurements reveal a spin Hall magnetoresistance (SMR) similar to that previously observed for antiferromagnetic (AFM) insulators/Pt heterostructures, indicating a broken-symmetry state with a characteristic AFM-like spin anisotropy with spin quantization axis transverse to an in-plane magnetic field. Notably, the observed SMR persists to fields where the AFM order may not exist. Our data also point to distinct energy scales, consistent with those found to govern the behaviors of various phases studied previously, including the enigmatic QSL phase. Our studies suggest that SMR could be an experimental technique complementary to others, such as neutron scattering and magnetic susceptibility measurements, capable of providing unique insights into exotic magnetism.

## Results

Our experimental technique is based on a bilayer heterostructure between $\alpha$-$RuCl_3$ and Pt, the most popular nonmagnetic material for SHE to detect spin properties in other materials. We employed SMR measurements (23,24) previously performed in a bilayer between Pt and a conventional (non-frustrated) magnetic insulator. In Pt (labeled "N" in Fig. 1A-1D), a current can generate spin accumulation on the surface (direct SHE: Fig.1A), and a spin current can be converted to a charge current (inverse SHE: Fig.1B).

One can use the current-induced spin accumulation on the Pt surface to interact with, and thus probe, the magnetic moments in a neighboring material, such as a magnetic insulator — in our case $\alpha$-$RuCl_3$ — interfaced with Pt. The backflow of the spin current (from the interface back to Pt) depends on the spin configuration in the neighboring material (labeled "M" in Figs.1C and 1D) due to an anisotropic scattering: the spin current prefers to be absorbed when the spins (labeled "$\sigma$") accumulated due to SHE at the Pt surface are perpendicular to the magnetic moments (labeled "m") in "M" (Fig.1C) and reflected when the spins are parallel (or antiparallel) to the moments (Fig.1D) (23-25). In the latter configuration (spins colinear with magnetic moments), the reflected spin current is converted to an additional charge current (on top of the one driven by the external voltage), leading to an increase of longitudinal conductivity in Pt. Such an anisotropic spin-charge conversion at the interface generates an apparent resistivity anisotropy. This phenomenon means that the electronic transport in Pt becomes dependent on the angle between the magnetic moment direction $\hat{X}$ (vector) in the magnetic insulator (M) and the current direction (Fig.1C and 1D), making the two longitudinal resistivities $\rho_{XX}$ (current flowing along the magnetic moment direction, Fig.1C) and $\rho_{YY}$ (current flowing perpendicular to the magnetic moment direction, Fig.1D) unequal (in general $\rho_{XX} > \rho_{YY}$ as explained above and observed previously (27,28)), where $\hat{Y}$ is perpendicular to $\hat{X}$.

Upon varying the angle ($\theta$) between the direction of the current ("$x$", thus the direction of current induced spin polarization $\sigma // y \perp x$) in the plane relative to the direction of the magnetic moment m (which may be controlled by an external magnetic field B), an oscillatory variation appears for both longitudinal and transverse resistivity components

$$\rho_{xx} = \frac{\rho_{XX}+\rho_{YY}}{2} + \frac{\rho_{XX}-\rho_{YY}}{2}\cos 2\theta, \qquad (1)$$

$$\rho_{xy} = \frac{\rho_{XX}-\rho_{YY}}{2}\sin 2\theta. \qquad (2)$$

Experimentally one often defines the direction of current with angle ($\alpha$) relative to the external in-plane magnetic field *B* (24). The expected oscillatory $\rho_{xx}$ and $\rho_{xy}$ as functions of $\alpha$ would be the same as in Eq. (1) and (2) for the "longitudinal anisotropy" case (where the moment m or $\hat{X}$ is parallel to *B* field, shown in Fig.1E), with $\alpha = \theta$. Whereas the expected oscillations would take the opposite (minus) signs for the "transverse anisotropy" case (where the moment m or $\hat{X}$ is transverse to *B* field, see Supplementary Note 1 for details), with $\alpha = \theta - \frac{\pi}{2}$. The resistivity oscillation can be conveniently utilized to characterize magnetic properties and spin-dependent phenomena --- even for a magnetic *insulator* with a small size --- by transport experiment, using proximal Pt.

We synthesized $\alpha$-$RuCl_3$ crystals via vapor transport from pure anhydrous powder (see *Materials and Methods* for details). The crystals were exfoliated into flakes with thicknesses of 50 nm (device #1, shown in Fig.2A) and 35 nm (device #2). We employed electron beam lithography with physical vapor deposition to define Pt Hall bar proximal to the flakes and the connecting Ti/Au electrodes (including contact pads on the silicon substrate). The mean roughness of the flake (measured in device #2) was about 0.03 nm (much less than the thickness of a 2D layer of $\alpha$-$RuCl_3$ which is 0.6 nm (4,8)) characterized by atomic force microscopy. Pt is on top of $RuCl_3$ for device #1, and $RuCl_3$ is on top of Pt for device #2. The magneto-transport measurements of the $\alpha$-$RuCl_3$/Pt Hall bars (on $SiO_2$/Si substrate) were performed by changing the strength and the direction of the magnetic field (the electrical probe configuration is shown in Fig.2B). The procedure for device fabrication, the transport experiment, and other characterizations are described in *Materials and Methods*.

Figure 2C shows clear angular oscillations of the transverse and longitudinal resistance as functions of the angle ($\alpha$, Fig.2B) between the current and the magnetic field of 18 T in the plane, with an approximate period of $\pi$. The oscillations in $R_{xy}$ and $R_{xx}$ are found to have phase shifts from $\cos 2\alpha$ by approximately $\pi/4$ and $\pi/2$ (behaving as $-\sin 2\alpha$ and $-\cos 2\alpha$), respectively (the respective dip positions are at ~45° and 0°). We have performed similar measurements on a Pt Hall bar without $\alpha$-$RuCl_3$ and did not observe any notable oscillations with a period $\pi$ (Fig.S2).

The observed oscillations resemble previously observed SMR for NiO/Pt bilayers, where the Néel vector, or in general the magnetic moments, in NiO (which is an antiferromagnet (AFM)), are transverse to the magnetic field direction (24,27). SMR oscillations in magnet/Pt bilayers have been observed in several magnetic insulator materials such as yttrium iron garnet (ferromagnetic (FM), with longitudinal anisotropy) (23), NiO (AFM) (27), and for $NiPS_3$ (van der Waals AFM (29)). SMR is also observed in bilayer with rare-earth transition ferrimagnet where orbital magnetic moment may dominate over spin magnetic moment (30). The spin properties are reflected over length scales of several nano-meters from the interface (23,27,29,31-33). We note even if the next layer below Pt have opposite (alternate) moments from the first layer (for example the case for NiO(111)), since they still have the same spin quantization axis that is transverse to the $B$ field, they contribute to the SMR phase just the same way. Our SMR amplitude ($\Delta\rho_{xy}/\rho$ ~$10^{-5}$) is similar to typical values observed previously (23,27,31). Additionally, when there is an out-of-plane component of the magnetic field $B_z$, which can appear with a slight misalignment of the field from the sample plane, the ordinary Hall effect (with ordinary Hall resistivity $\rho_{Hall}$) can give rise to a $\rho_{Hall}B_z$ term in the transverse resistivity and thus an additional resistance oscillation with a period of $2\pi$ (e.g., indicated as $\Delta R_{xy}^{2\pi}$ in Fig.2C). Accordingly, $2\pi$-periodic oscillations in the transverse resistivity may appear via finite transverse susceptibility (34), but should be of even smaller magnitude and are unlikely to make an appreciable contribution.

The magnetic field dependence of the SMR signal (angular oscillation) for device #1 is shown in Fig.3A (representative traces) and Fig.3B (plotted in interpolated color scale, with all traces shown in Fig.S3). To characterize the amplitude and the phase of the angular oscillations, the SMR data were fitted to

$$\Delta R_{xy}^{\pi}\cos(2(\alpha-\alpha_{SMR})+\pi)+\Delta R_{xy}^{2\pi}\cos(\alpha+A)+R_0+C\alpha,$$

where $\Delta R_{xy}^{\pi}$, $\alpha_{SMR}$, $\Delta R_{xy}^{2\pi}$, $A$, $R_0$ (the offset), and $C$ (the linear drift slope) are free-fitting parameters (see Supplementary Note 2 for details). For device #1, we observed a phase $\alpha_{SMR}$ close to that (45°) expected for SMR in AFM/Pt (such as NiO/Pt) over the entire range of fields studied (~ 1.5 T to 18 T). On the other hand, the amplitude grows rapidly with the field up to about $B_C$ ~ 7 T. Above 7 T, the amplitude no longer increases appreciably and rather saturates (Fig.3C). Those trends are more or less similarly observed on another device (#2) shown in Figs.3C and 3D (though the phase $\alpha_{SMR}$ is notably larger for fields < 7 T).

We explore the energy scales by the temperature dependence of the SMR measured in $\alpha$-$RuCl_3$/Pt (device #2). An overarching feature is that the SMR oscillation amplitude rapidly decreases with increasing temperature (blue shaded region in Fig. 4A) until becoming difficult to discern (at least for lower B fields) given the noise level of the measurement (Fig.4A). This feature is pronounced below $B_C$ where a linear fit to the temperature-dependent amplitude on the low temperature side (blue region in Fig.4A) shows the zero-amplitude intercepts at $T_C$ = 6.6 ± 0.5 K for $B$ = 5.0 T and $T_C$ = 8.2 ± 0.5 K for $B$ = 4.0 T (overall average of $T_C$ = 7.9 K, over four fitted lines for 4.0-8.0 T in Fig.4A). For higher B fields, the SMR remains observable at higher temperatures (see green shaded region in Fig. 4A, up to the highest temperature studied in our experiment).

**Discussion**

We point out that SMR signal along with associated "transverse-anisotropy" or "AFM-like" angular oscillations (Fig. 1F) are expected to be present as long as (i) there exists a preferred spin quantization axis $\pm\hat{X}$ for the magnetic moments (with spin directions $\hat{S}$) in the (magnetic) insulator interfaced with Pt and (ii) the quantization axis is mostly perpendicular to the field (with axis $\hat{B}$)

$$\hat{S} \,//\, \pm\hat{X},\ \angle(\hat{S},\hat{B}) \approx \pm\pi/2. \quad (3)$$

Note the contribution to SMR due to the anisotropic interfacial spin-charge conversion mechanism described above may not change even if $\hat{S}$ reverses the sign, still giving apparently the same SMR signal. This means neither a magnetic order nor a spatial periodicity in magnetic structure is strictly needed to generate the AFM-like SMR and associated resistance oscillations. It suffices to have a *B*-transverse anisotropy axis ($\angle(\hat{S},\hat{B}) \approx \pm\pi/2$, regardless of whether any magnetic order is present or not).

Importantly, such a spin configuration does not need to be static. A state described by (3) is exemplified by the inset of Fig. 5 (with a brown background), where each spin has a substantial expectation value for the component perpendicular to the applied field (see Supplementary Note 3 for further discussions). A fluctuating zigzag order (where there could be substantial fluctuating moments largely transverse to the field, even if there are static moments longitudinal to the field ) is suggested by the torque measurement at fields higher than $B_C$ (35). The magnitude of the SMR, $\Delta R_{xy}^{\pi}$, is sensitive to the anisotropy of the direction of the spins in the material adjacent to the Pt surface, and how the anisotropy direction co-rotates with the B-field relative to the direction of electrical current $\overrightarrow{J_{Pt}}$ (Figs. 1E,F) – with a higher magnitude of SMR indicating a greater degree of this anisotropy. Since $\overrightarrow{J_{Pt}}$ is an axial vector, the response of the SMR is always sensitive to an axial $C_2$ (e.g., *B*-transverse) anisotropy. Note that other alignment of the spins in the adjacent material incompatible to an overall $C_2$ anisotropy, such as random spin orientations, or a $C_3$ anisotropy, will lead to a complete cancellation of the SMR signal. The phase $\alpha_{SMR}$, on the other hand, is dependent on the relative orientation between the axis of the magnetic field and the spins in the interfacial material (with $\alpha_{SMR} = 3\pi/4$ for B-longitudinal spins and $\pi/4$ for B-transverse spins, respectively, as depicted in Fig. 1E,F).

Figure 5 shows a cartoon phase diagram reflecting our interpretation of the spin configurations, showing the direction and the strength of the spin anisotropy in $\alpha$-$RuCl_3$ based on our data. At the zero-field and low temperatures $\alpha$-$RuCl_3$ is a zigzag AFM and has three types (equally likely) of zig-zag AFM domains, preserving an overall $C_3$ symmetry (14). Up to the modest fields of ~1 T, the spins and the domains do not vary with the external field, maintaining the overall $C_3$ symmetry, and hence giving no SMR signal (Fig. 3B). Between ~1.5 T (36) and ~7 T, while the spins within each domain remain locked to the local zigzag direction, the percentages of different types of domains redistribute so that the domains whose spins are more [less] transverse to the *B* field is more [less] favored (thus would grow [shrink]) according to the magnetic field strength and orientation. Therefore, the spins of the whole system on average can now develop a $C_2$ magnetic anisotropy toward to be more transverse to the *B* field (even the spins within each domain may still be pinned to the local zigzag crystal orientation), as suggested from the bulk measurements (11, 14). It is well known that the zig-zag magnetic order starts to attenuate with increasing *B*-field and the suppressed magnetic order leads to a larger fraction of fluctuating moments. The increasing SMR signal is accounted for by a larger fraction of spins more transverse to *B* by the expansion of corresponding types of domains, as well as the increase of fluctuating moments preferring to align transversely to *B*-field. The SMR phase represents the net *B*-transverse anisotropy, that can vary at low B field due to the imperfect domain distribution, and also varies between samples (Fig.2D). The signal stops increasing approximately at the field of 7 T, which broadly corresponds to where the zigzag ordered phase ends (19, 37).

Notably, the temperature dependance of the magnitude of the SMR signal suggests an order-parameter that gets suppressed at $T_C$ ~ 8 K at 4 T and 6.6 K at 5 T, seemingly closely following the order parameter for the zig-zag state. At higher temperatures, the increased thermal fluctuations lead to the scrambling of the net spin anisotropy. The temperature regime between 10 – 100 K is known to host a thermally driven Kitaev paramagnetic phase shown in both neutron scattering (38,39) and Raman (5, 6) experiments. However, this phase is $C_3$ symmetric and expectedly leads to a net zero SMR signal (Fig.5).

At higher fields >10 T (but much lower than the saturation field of >40 T [10,21,40,41]), $\alpha$-$RuCl_3$ is believed to enter a partially field polarized regime. Our $\alpha_{SMR}$ approaches $\pi/4$ (Fig 4D), indicating spins mostly transverse to the magnetic field (even more so at higher fields and higher temperatures, indicated by $\alpha_{SMR}$ being closer to $\pi/4$). Prior magnetization measurements (10,21,41) confirmed static spins are mostly longitudinal to in-plane *B* field. These suggest significantly strong quantum fluctuations, with a spin anisotropy axis largely perpendicular to the applied field, even at field as high as 18 T. Electron spin resonance experiment (42) suggested the dominant excitation intensity is for THz B-field polarized perpendicular to the field, that may corroborate with the dynamical spin existing in transverse direction. The results may also corroborate the torque magnetometry results (40) which suggested an extended critical phase with magnetic anisotropy with respect to the polar angle of the B field robust to very high fields and also significantly above the low field critical temperature of ~ 8 K.

For high fields > 8 T (e.g., at 11.5 T, 13.5 T, and 14.5 T studied in Fig. 4A), upon increasing temperature to a critical temperature ~ 6 K the SMR intensity displays an order-parameter-like drop (to finite but nonzero values, Fig. 4A). This temperature scale for the initial drop is not so different from the temperature scale for the vanishing of SMR intensity at lower fields, overall representing an important anisotropy energy scale for $\alpha$-$RuCl_3$.

The green shaded region in Fig. 4A seems to indicate the presence of another and larger spin anisotropy related energy scale (corresponding to a temperature scale much higher than ~ 6 K) at higher fields. The observation that finite SMR intensity can persist to even higher temperatures (after the initial drop) indicates that the increased spin fluctuations at such large in-plane magnetic fields (> 10 T) may help the spins to align more transverse to $\vec{B}$. This behaviour may correspond to a phase in which previous Raman experiments (43) have reported a larger spin gap (possibly linked to anisotropy) with increasing field.

Our measurements mark a significant advancement bringing together the fields of quantum materials and spintronics. Spin-sensitive probes are rare and so are electrical probes of insulating quantum magnets, including spin liquid candidates (44) like $\alpha$-$RuCl_3$ despite that such measurements are highly desired (12). Here we have showcased a transport device that can be adopted to characterize the anisotropy (either for static and fluctuating spins) at the wide field and temperature range including finite-field phase of $\alpha$-$RuCl_3$. Our approaches allow using electric transport to probe insulating materials, even for atomically thin samples, which allows studying dimensional crossover and interlayer coupling. In addition, the spin-charge coupling utilized in our experiment may be useful to electrically control magnetic states, although understanding of the detailed coupling mechanisms between strongly spin-orbit coupled materials and spin liquids is still at a very early stage. Finally, we believe that our method can be further combined with, or adapted to, other experimental techniques such as scanning probe microscopies and high-frequency transport in similar heterostructures (to probe quantities which pertain to the spatial and dynamic correlation between spins), and is complementary to other techniques such as Raman and neutron scattering. As exemplified in the studies of many other prominent quantum materials (from high Tc superconductors to topological insulators), different signatures of novel states of matter are often detected by multiple probes, leading to a fuller picture of the system.

**Materials and Methods**

## 1. Crystal synthesis

Single-crystal $RuCl_3$ was synthesized via chemical vapor transport from pure anhydrous powder. Commercial $\alpha$-$RuCl_3$ powder was sealed under vacuum in a quartz ampoule and heated to 1060 °C at a rate of 1.6 °C/min. The ampoule was held at this temperature for 12 hours before being cooled to 600 °C at a rate of 6 °C/hr, at which point the furnace was allowed to cool back to room temperature. The crystals grew as shiny black plates, typically around 0.5 cm x 0.5 cm in lateral size and around 0.5 mm in thickness, though some larger crystals were also obtained. Samples were characterized by X-ray diffraction to determine phase purity. Samples were also characterized by magnetic susceptibility measurements. All the crystals were measured and a single peak in the susceptibility was observed at approximately 7.85 K (Fig.S4), which has been shown to be evidence of low stacking fault density (8).

## 2. Procedure of device fabrication, transport experiment and characterization

For device #1, the $\alpha$-$RuCl_3$ flakes were mechanically cleaved from the crystals on silicon substrate (Si with 285-nm-thick $SiO_2$).The standard *e*-beam lithography process was applied to pattern the platinum Hall bar, titanium/gold (depositing 5-nm-thick Ti on $SiO_2$ to improve the adhesion followed by 60-nm-thick Au) electrical leads and pads. The optical micrograph of the device is shown in Fig.1E. For device #2, we first prepared a 3-nm-thick Pt film on silicon substrate and then $\alpha$-$RuCl_3$ was mechanically cleaved on the Pt film. After the Au pads are formed on Pt (at the areas without the $\alpha$-$RuCl_3$ flake), the $\alpha$-$RuCl_3$/Pt bilayer is patterned to a Hall bar by *e*-beam lithography and Ar ion-milling. Thickness and the roughness were characterized by atomic force microscopy. The thicknesses of the $\alpha$-$RuCl_3$ flakes were 50 nm for device #1 and 35 nm for device #2. The mean roughness of the flake for device #2 was about 0.03 nm. The $\alpha$-$RuCl_3$ flakes were characterized by Raman microscopy after the *e*-beam process and did not show noticeable degradation. Further, we performed scanning transmission electron microscopy for another $\alpha$-$RuCl_3$/Pt bilayer ($\alpha$-$RuCl_3$ thickness is ~ 8 nm), which indicated clear crystalline $\alpha$-$RuCl_3$ without significant disorder at the interface (Fig.S5). The magneto resistance measurement of Pt/$\alpha$-$RuCl_3$ Hall bars were performed (the electrical probe configuration is shown in Fig.2B) in a cryostat with superconducting magnet and a two-axis sample rotation probe. The rotation plane is carefully calibrated using a Hall sensor and the perpendicular component of the field is minimized. Because of the aspect ratio of device #1, we did not carry out antisymmetrization and just use the $R_{xy}$ value itself. For device #2, the aspect ratio is smaller than device #1 and we carried out antisymmetrization by subtracting the $R_{xy}$ data with positive $B$ field from negative $B$ field and divided by 2.

## Acknowledgments

We thank Q. Niu, D. Xiao, G. Bauer, and X. Wang for helpful discussions. This work was supported in part by AIMR and its “fusion” research program, under World Premier International Research Center Initiative (WPI), the Ministry of Education, Culture, Sports, Science and Technology (MEXT), Japan, and by Grant-in-Aid for Scientific Research, JSPS KAKENHI (Grant Numbers 22H01896, 22H00278, 21K18590, 20H04623, 20K14399, 18H03858 and 19K03736), by Center for Science and Innovation in Spintronics, by GIMRT Program (High Field Laboratory for Superconducting Materials) of the Institute for Materials Research, by Center for Integrated NanoTechnology Support, and by the electron microscopy center, Tohoku University. The research by S.O. and P.L. was supported by the Scientific Discovery through Advanced Computing (SciDAC) program funded by the US Department of Energy (DOE), Office of Science, Advanced Scientific Computing Research. N.M. was supported by the DOE, Office of Science, Basic Energy Sciences, Materials Sciences and Engineering Division. M.C., S.E.N., and D.M. acknowledge support from National Science Foundation (NSF grant number DMR-1808964). S.O. (later stage of his research), A.B. and Y.P.C. also acknowledge support by Quantum Science Center (QSC), a US DOE National Quantum Information Science Research Center.

**Author Contributions**

H.I., K.M. and Y.P.C. conceived the project. M.C., D.M. and S.E.N. grew and characterized the single crystals of $\alpha$-$RuCl_3$. H.I. fabricated the device. H. I. performed transport measurement and analyzed the data with the help of K.M. S.O. P.L. N.M., and A.B. contributed to the interpretation of the results. H.I. and Y.P.C. wrote the first draft and the manuscript was finalized with input from S.O., A.B., S. E. N. and all other coauthors.

**Competing Interest Statement:** The authors declare no competing interests.

## Figures and Tables

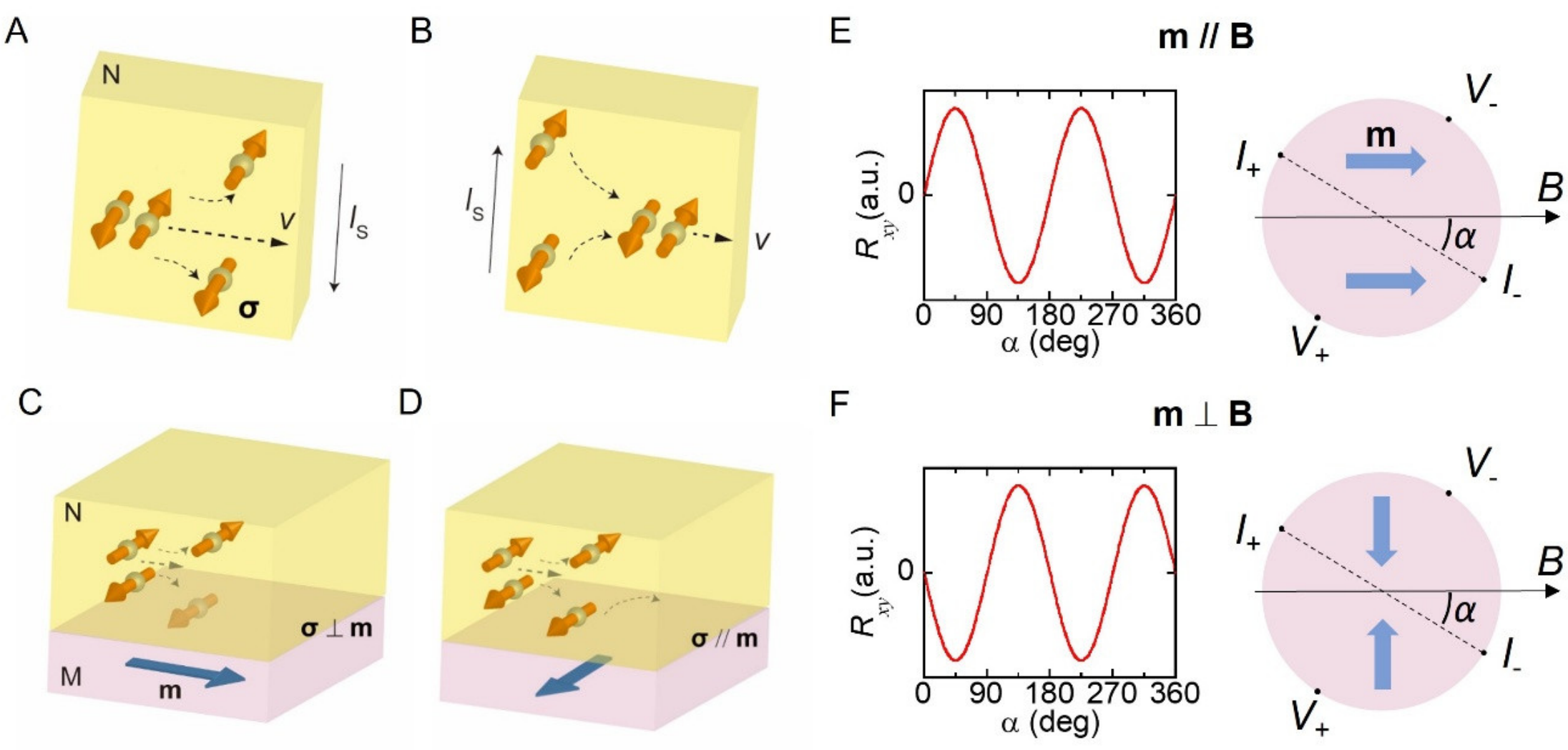

**Figure 1. Spin-Hall magnetoresistance (SMR) as spin sensitive transport.** (A) Direct spin Hall effect in a nonmagnetic spin-Hall metal N (such as Pt used in our work). The charge current (indicated by velocity *v*) induces spin current *I*s, leading to spin (σ) accumulation at the surface. (B) The inverse spin Hall effect converts spin current to electric charge current. (C,D) Spin Hall effect in a magnet(M)/ N bilayer. When current flows in N, the spin Hall effect generates spin polarization (σ) accumulation at the interface. The spin-dependent interaction between σ and the magnetic moment (**m**) in the magnetic material (M) depends on their relative angle. The different interactions between the transverse (C) and parallel (D) cases cause different efficiencies for spins to be reflected from the interface back to N and converted to charge current. The longitudinal resistivities are therefore different (generally higher in C than in D), giving rise to SMR (including an apparent transverse resistivity for a generic current direction). (E,F) Expected angular oscillation of the transverse resistance for the two cases where **m** (indicated by blue arrows) aligns parallel (E) or transverse (F) to the external field *B* direction (with angle α from the electrical current direction).

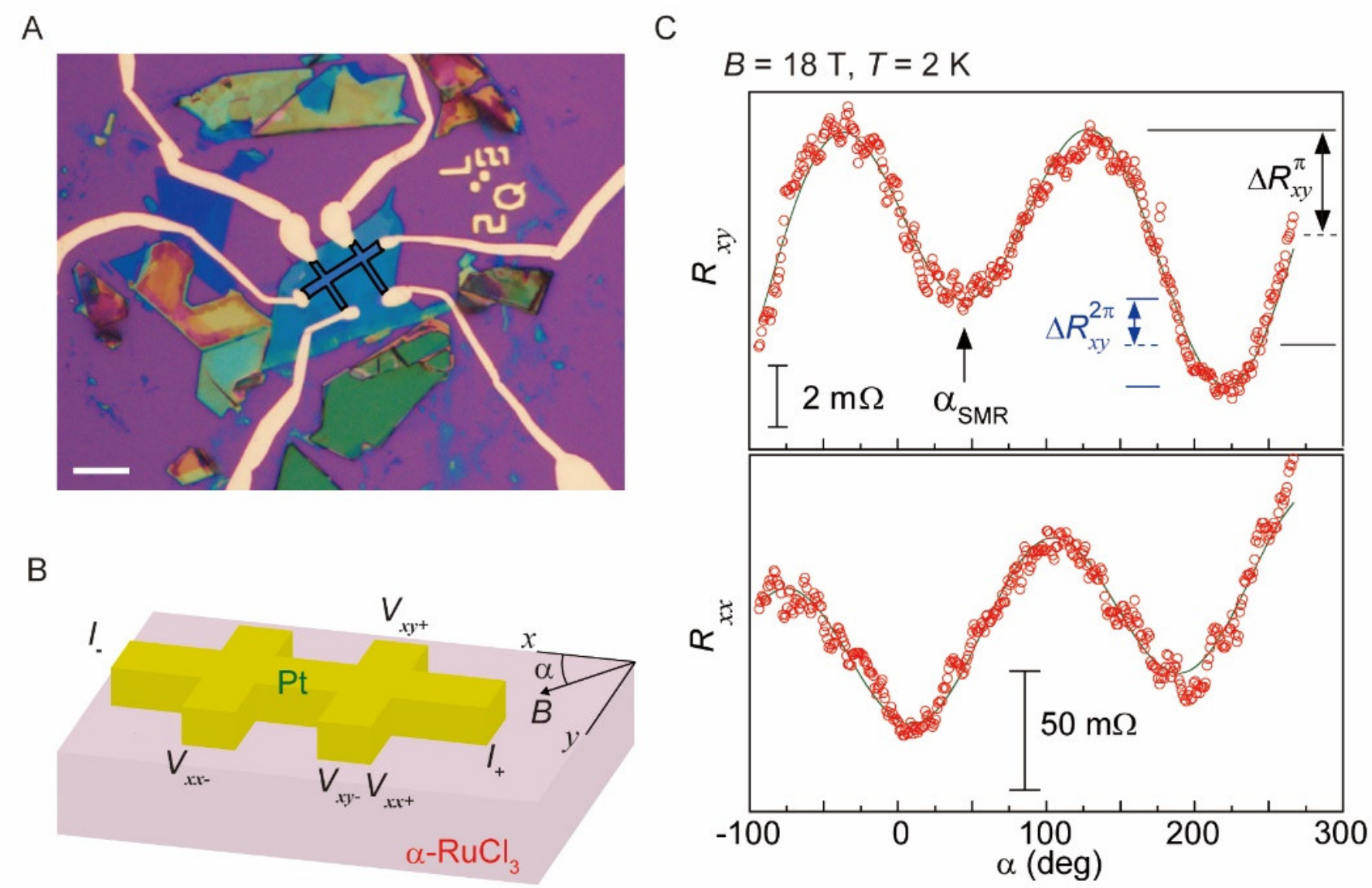

**Figure 2. Measured spin-Hall magnetoresistance (SMR).** (A) Optical micrograph of an α-RuCl₃/Pt heterostructure (device #1). Scale bar is 10 μm. The 3-nm-thick Pt Hall bar is fabricated on the flat surface of a cleaved α-$RuCl_3$ flake (whose thickness is 50 nm). The black line indicates the boundary of the Hall bar. (B) Schematic of device and electrical transport measurement. The in-plane angle $\alpha$ of the magnetic field (*B*) is defined from the direction (*x*) of current flow. (C) Oscillation of transverse (longitudinal) resistances $R_{xy}$ ($R_{xx}$) as functions of α measured with the in-plane magnetic field *B* of 18 T and at *T* = 2 K. $\Delta R_{xy}^{2\pi(\pi)}$ indicate the amplitudes of fitted oscillation components with a period of 2π (π). $\alpha_{SMR}$ denotes the first positive angle position of the local minimum in the $R_{xy}$ oscillation component with the period π.

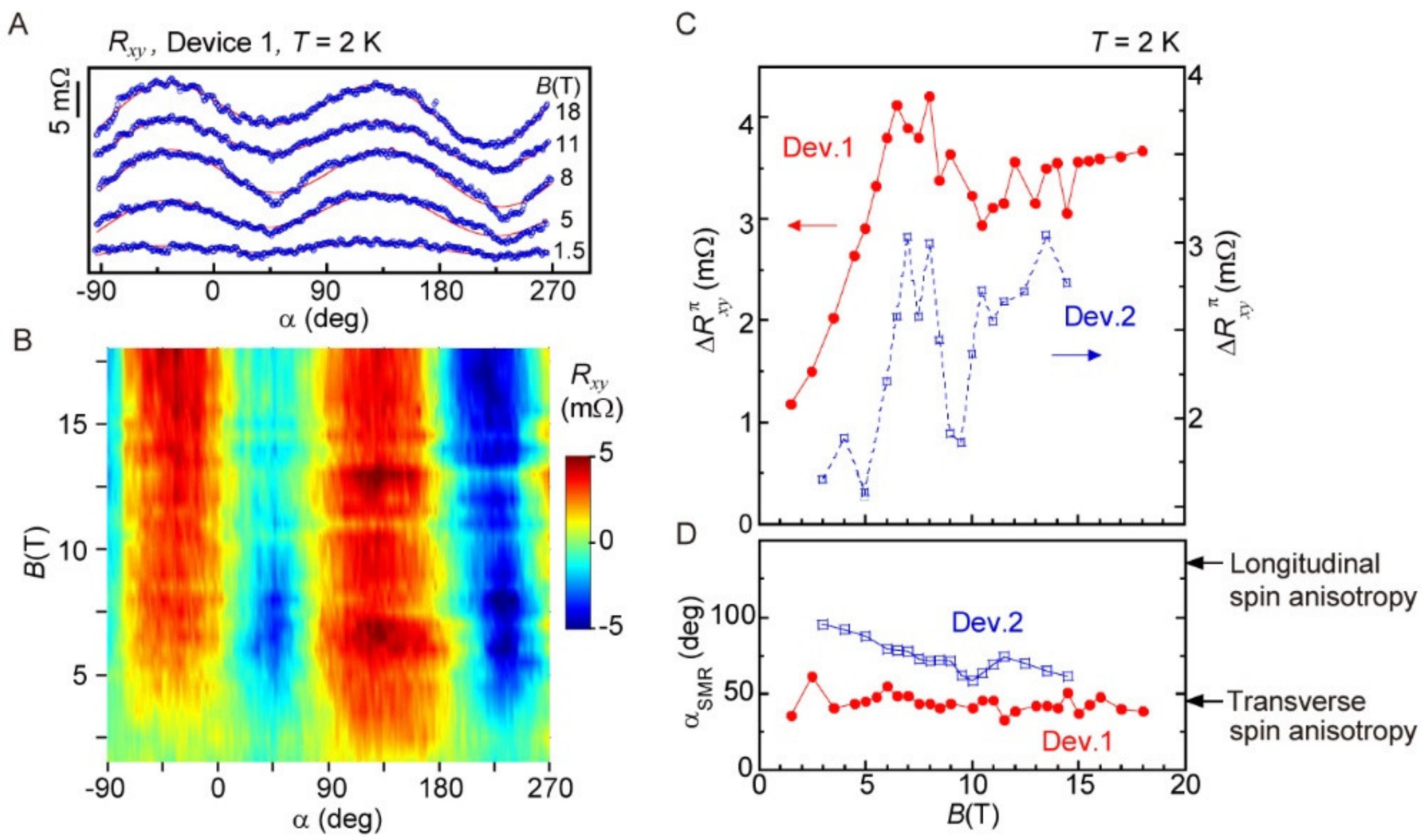

**Figure 3. Magnetic field evolution of the angular oscillation of SMR.** (A) The angular oscillation in the transverse resistance ($R_{xy}$) at $T$ = 2 K under representative magnetic fields of 18 T, 11 T, 8 T, 5 T, and 1.5 T. The solid red lines are obtained by fitting (see text for details). The curves are shifted vertically for clarity. (B) The magnetic field ($B$) evolution of the angle ($\alpha$)-dependent $R_{xy}$ (with interpolation between traces measured at adjacent $B$ fields) shown as a color plot (scale bar on the right). A constant background is fitted and subtracted from each trace for clarity. (C,D) The magnetic field evolution of the amplitude (C) and the phase (D) of the $\pi$-periodic oscillation in $R_{xy}$ at $T$ = 2 K for the device #1 (red closed circle, labeled as Dev. 1) and for device #2 (blue open rectangle, Dev. 2). The expected $\alpha_{SMR}$ for the case of spin anisotropy longitudinal and transverse to the $B$ field are indicated by arrows.

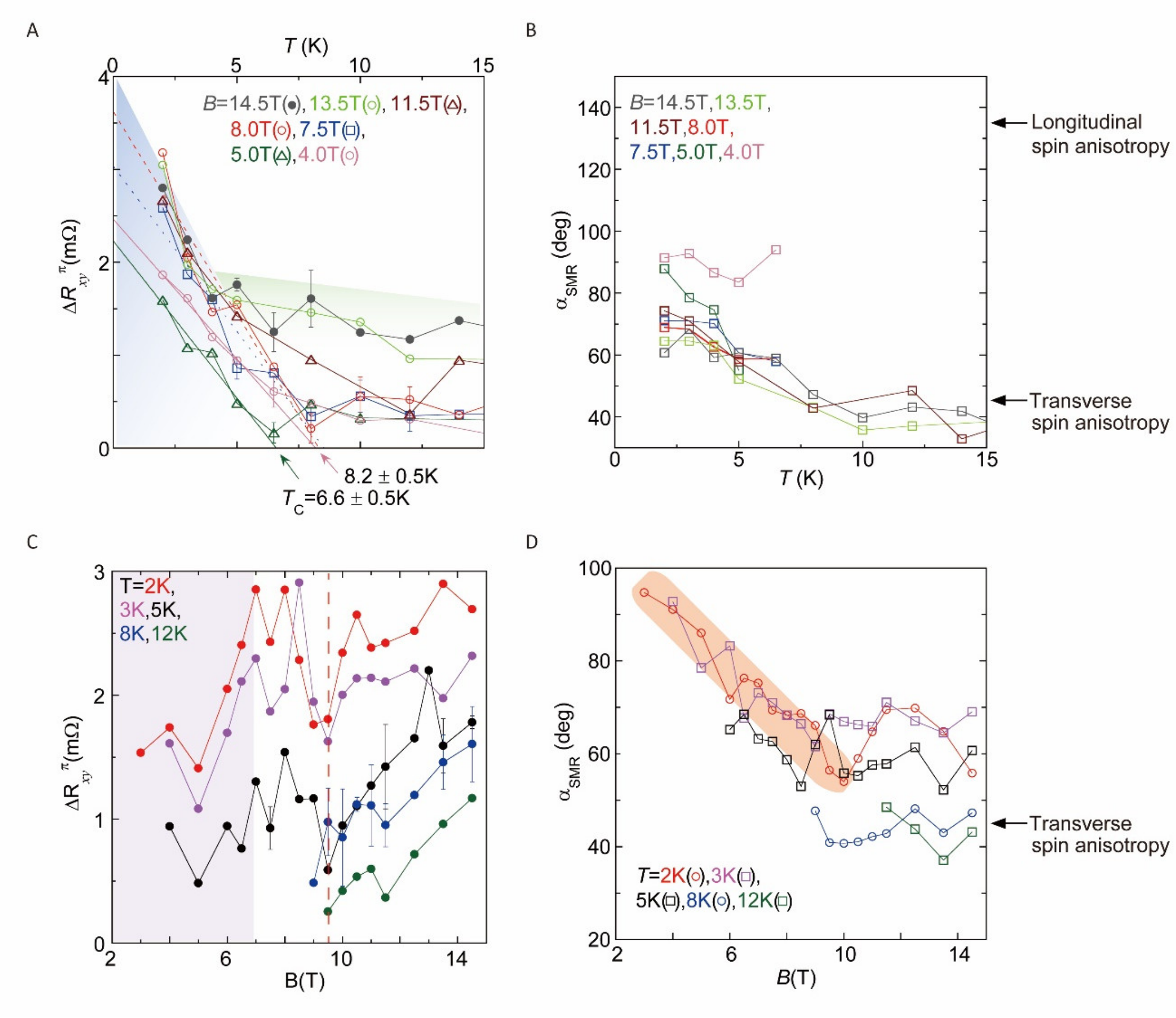

**Figure 4. SMR at elevated temperatures.** (A) Temperature evolution of the SMR oscillation amplitude $\Delta R_{xy}^{\pi}$ measured for device #2 under the in-plane fields of 14.5 T (gray), 13.5 T (light green), 11.5 T (brown), 8.0 T (red), 7.5 T (blue), 5.0 T (green), and 4.0 T (pink). The linear fits show the interceptions to $\Delta R_{xy}^{\pi}$ = 0 at the temperatures of 6.6 K and 8.2 K for *B* = 4.0 T and *B* = 5.0 T, respectively (shown by green and pink arrows) The region for the rapid linear decrease with temperature is marked by the shaded blue area. The other region (with slower decrease with temperature) is marked by the shaded green area. Uncertainty bars shown are representative. (B) The temperature evolution of the phase ($\alpha_{SMR}$) of the SMR oscillation for the same set of magnetic fields as in (A). The expected phases for SMR oscillation for the cases of longitudinal and transverse spin anisotropies (Fig. 1E and 1F respectively) are indicated by arrows. (C) The field evolution of the SMR amplitude for the temperature of 2 K (red), 3 K (violet), 5 K (black), 8 K (blue) and 12 K (green). The purple box indicates the expected magnetic field range of the AFM phase and the red dotted line is shown at *B* = 9.5 T (where a possible dip in the SMR signal amplitude is noted, though more work is needed to further investigate this feature). Uncertainty bars shown are representative. (D) The field evolution of the SMR phase for the same set of temperatures as in (C). The region where the phase rapidly changes is highlighted by the shaded orange rectangle. Typical uncertainty of $\alpha_{SMR}$ is no more than 10°.

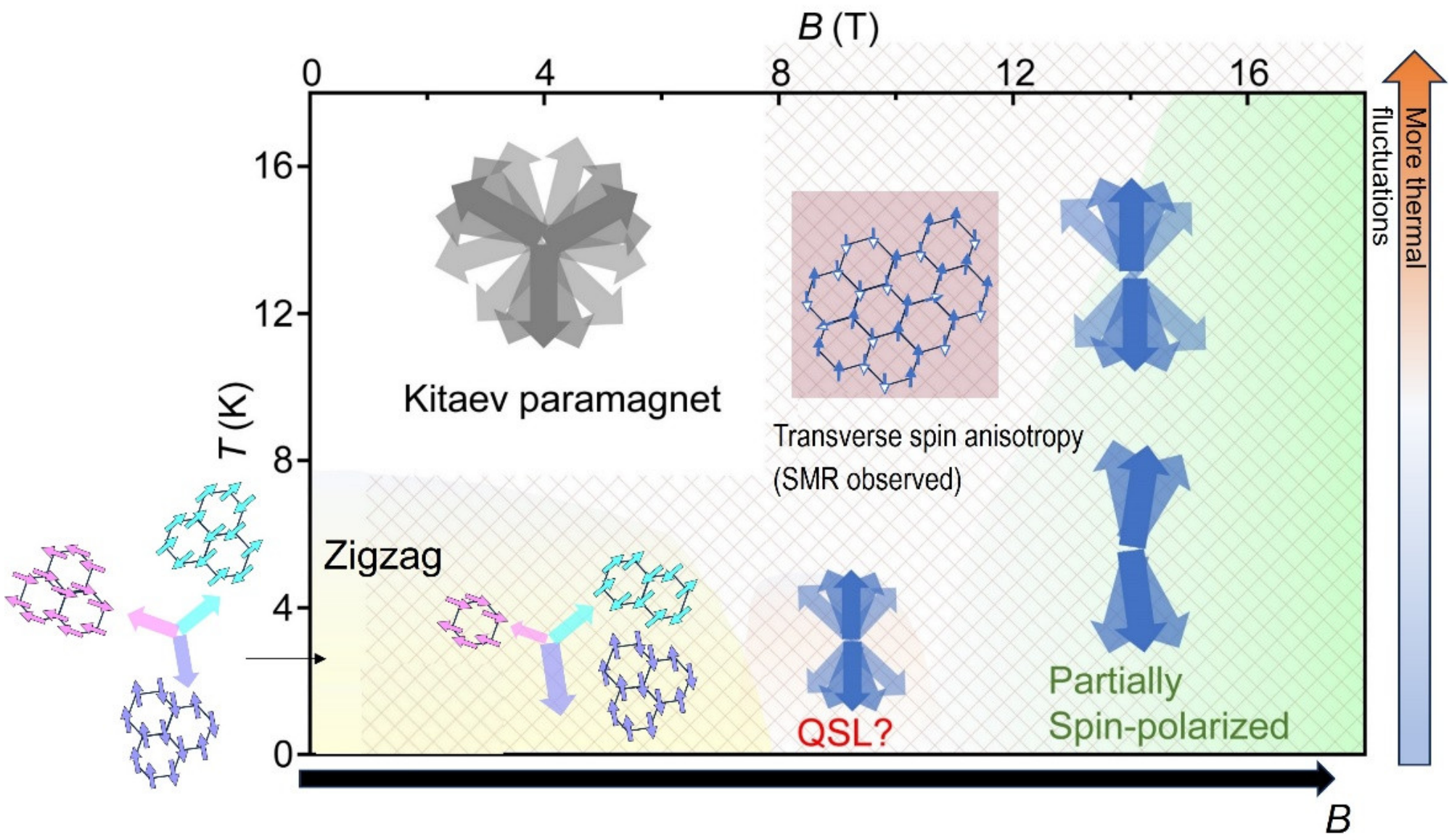

**Figure 5. An experimental phase diagram of α-$RuCl_3$.** Hatched region indicates SMR observed, indicating an underlying spin anisotropy (found to be largely transverse to the *B* field, regardless of static or not). Yellow, red, and green shaded regions show previous experiments on bulk crystals have determined as the Zigzag AFM, possible QSL phase, and "partially spin-polarized phase", respectively. Blue (in QSL and partially polarized regions) and gray (in paramagnet case) wide arrows indicate spins when the Zigzag AFM order is no longer expected. The inset with a brown background shows an example of proposed possible snapshot of local spin (S) configuration relevant for SMR in α-$RuCl_3$ under external in-plane (here in horizontal direction) magnetic field B (see Fig. S6 for details).

# Supplementary Materials for

# Spin sensitive transport in $RuCl_3$/Pt heterostructures: revealing a field-transverse spin anisotropy in a spin liquid candidate

H. Idzuchi[1-3*], M. Kimata[4], S. Okamoto[5,6], P. Laurell[7], N. Mohanta[5], M. Cothrine[8], S. E. Nagler[9], D. Mandrus[8], A. Banerjee[2,6] and Y. P. Chen[1,2,6,10,11*]

[1] *WPI Advanced Institute for Materials Research (AIMR) and Center for Science and Innovation in Spintronics (CSIS), Tohoku University, Sendai 980-8577, Japan.*
[2] *Purdue Quantum Science and Engineering Institute and Department of Physics and Astronomy, Purdue University, West Lafayette, Indiana 47907, USA.*
[3] *Department of Physics, The University of Tokyo, Bunkyo-ku, Tokyo 113-0033, Japan.*
[4] *Institute for Materials Research (IMR), Tohoku University, Sendai 980-8577, Japan.*
[5] *Materials Science and Technology Division, Oak Ridge National Laboratory, Oak Ridge, Tennessee, 37831 USA.*
[6] *The Quantum Science Center, Oak Ridge National Laboratory, Tennessee, 37831 USA*
[7] *Computational Sciences and Engineering Division, Oak Ridge National Laboratory, Oak Ridge, Tennessee 37831, USA.*
[8] *Department of Materials Science and Engineering, University of Tennessee, Knoxville, Tennessee 37996, USA.*
[9] *Neutron Scattering Division, Oak Ridge National Laboratory, Oak Ridge, Tennessee 37831, USA.*
[10] *School of Electrical and Computer Engineering and Birck Nanotechnology Center, Purdue University, West Lafayette, Indiana 47907, USA.*
[11] *Institute of Physics and Astronomy and Villum Centers for Dirac Materials and for Hybrid Quantum Materials and Devices, Aarhus University, 8000 Aarhus-C, Denmark.*

***Email:** idzuchi@g.ecc.u-tokyo.ac.jp (H. Idzuchi), arnabb@purdue.edu (A. Banerjee), yongchen@tohoku.ac.jp (Y. P. Chen)

**This PDF file includes:**

Supporting text
Figures S1 to S6
SI References

**Supplementary Information Text**

**Supplementary Note 1. Relation between resistance anisotropy (unequal $R_{XX}$ and $R_{YY}$) and angular oscillation in spin-Hall magneto resistance (SMR) for FM and AFM cases**

We consider the effect of anisotropic longitudinal resistivity. As described in text and depicted in Figs.1C-1D, the longitudinal resistance depends on the angle between the direction of magnetic moment ($X$) and spin accumulation induced perpendicular to the current direction ($x$). We define the $X$-$Y$ frame with $X$ parallel to the magnetic moment (which for FM is along $B$ direction, and for AFM is perpendicular to $B$ direction) and $Y$ perpendicular to $X$ (Fig.S1A). We assume that relevant magnetic moment and thus the ($X$,$Y$) frame corotate with the magnetic field. We define $x$-$y$ frame relative to the Hall bar (Fig.S1A), where the $x$ axis is parallel to the current direction (Fig.1F) and $y$ is perpendicular to $x$. The angle between $x$ (current) and $X$ (moment) is $\theta$. A conversion between frame ($X$,$Y$) relative to magnetic moment direction and frame ($x$,$y$) relative to Hall bar can be given as

$$\begin{pmatrix} \cos\theta & -\sin\theta \\ \sin\theta & \cos\theta \end{pmatrix} \begin{pmatrix} \rho_{XX} & 0 \\ 0 & \rho_{YY} \end{pmatrix} \begin{pmatrix} \cos\theta & \sin\theta \\ -\sin\theta & \cos\theta \end{pmatrix} = \begin{pmatrix} \rho_{xx} & \rho_{xy} \\ \rho_{yx} & \rho_{yy} \end{pmatrix}.$$

Here the first subscript for a resistivity component ($\rho$) refers to the current flow direction and the second subscript refers to the voltage probe direction used in the measurement (for example with current flowing along the $x$ direction, $\rho_{xx}$ is the longitudinal resistivity and $\rho_{xy}$ is the transverse or "Hall" resistivity shown in Fig.1E, whereas $\rho_{XX}$ ($\rho_{YY}$) is the longitudinal resistivity with current flowing parallel (perpendicular) to the magnetic moment direction). This means the longitudinal resistivity measured in $x$ direction (channel of Hall bar, see Fig.1E, with the magnetic moment at angle $\theta$) is given by

$$\rho_{xx} = \rho_{XX}\cos^2\theta + \rho_{YY}\sin^2\theta = \frac{\rho_{XX}+\rho_{YY}}{2} + \frac{\rho_{XX}-\rho_{YY}}{2}(\cos^2\theta - \sin^2\theta) = \frac{\rho_{XX}+\rho_{YY}}{2} + \frac{\rho_{XX}-\rho_{YY}}{2}\cos 2\theta,$$

and transverse resistivity given by $\rho_{xy} = (\rho_{XX} - \rho_{YY})\sin\theta\cos\theta = \frac{1}{2}(\rho_{XX} - \rho_{YY})\sin 2\theta$. Therefore, longitudinal (transverse) resistance shows angular oscillation with $\cos 2\theta$ ($\sin 2\theta$) in SMR (noting in general $\rho_{XX} - \rho_{YY} > 0$ as explained in the main text due to simultaneous action of direct spin Hall effect and inverse spin Hall effect with anisotropic interfacial spin transport coefficient (1)).

We note experimentally one often defines the angle ($\alpha$) between current direction $x$ and magnetic field $B$ (which is the same direction as $X$ for FM, thus $\alpha = \theta$, Fig. S1A). The experimentally measured SMR (as functions of $\alpha$) would be given by the same equations above with $\theta$ replaced by $\alpha$. For AFM case, we assume the magnetic moment direction $X$ is perpendicular to and co-rotates with the magnetic field $B$ in-plane (Fig.S1B). This gives an additional angle of $\pi/2$ between magnetic moment direction $X$ and magnetic field ($B$) direction hence $\theta=\alpha+\pi/2$. The SMR formulae for AFM now read

$$\rho_{xx} = \frac{\rho_{XX}+\rho_{YY}}{2} - \frac{\rho_{XX}-\rho_{YY}}{2}\cos 2\alpha,$$

$$\rho_{xy} = -\frac{\rho_{XX}-\rho_{YY}}{2}\sin 2\alpha.$$

**Supplementary Note 2. Fitting of SMR signal**

In this work, SMR component is extracted by fitting the measured transverse resistance $R_{xy}(\alpha)$ to the following formula

$$\Delta R_{xy}^{\pi}\cos(2(\alpha-\alpha_{SMR})+\pi)+\Delta R_{xy}^{2\pi}\cos(\alpha+A)+R_0+C\alpha+D\alpha^2 \quad \text{(S1)}$$

where the SMR term is $\Delta R_{xy}^{\pi}\cos(2(\alpha-\alpha_{SMR})+\pi)$ and a misalignment of the field (tilted out of plane) gives $\Delta R_{xy}^{2\pi}\cos(\alpha+A)$. In fitting results from the same set of the measurement, we used the same value of $A$.

(i) For data at 2 K (Figs.2 and 3), we set $D = 0$: the equation is now:

$$\Delta R_{xy}^{\pi}\cos(2(\alpha-\alpha_{SMR})+\pi)+\Delta R_{xy}^{2\pi}\cos(\alpha+A)+R_0+C\alpha \quad \text{(S2)}$$

(Step i-i) We first determine $A$. The term $\Delta R_{xy}^{2\pi}\cos(\alpha+A)$, is found larger for high field. We analyze the data at $T$=2 K under high field and fixed the coefficient $A$.

(Step i-ii) The data are fitted to (S2) with free parameters of $R_0$, $\Delta R_{xy}^{\pi}$, $\alpha_{SMR}$, $C$, and $\Delta R_{xy}^{2\pi}$ ($A$ is already determined in Step (i-i)).

**Example**: fitting to the top panel data of Fig.2C to (S2) with $A$ = 30.23° gives:

- $R_0$ = 64.5 ± 0.04 mΩ,
- $\Delta R_{xy}^{\pi}$ = 3.79 ± 0.04 mΩ,
- $\alpha_{SMR}$ = 45.4 ± 0.3°,
- $C$ = 0.0069 ± 0.0004 mΩ/deg,
- $\Delta R_{xy}^{2\pi}$ = 2.10 ± 0.05 mΩ.

We expect (see Note 1 above) that $\alpha_{SMR}$ = 45° for AFM and 135° for FM.

(ii) For the temperature dependence data (Fig.4), we use (S1) instead of (S2). This means we include the full quadratic drift term $R_0+C\alpha+D\alpha^2$.

(Step ii-i) We first determine both $\Delta R_{xy}^{2\pi}$ and $A$. The term

$$\Delta R_{xy}^{2\pi}\cos(\alpha+A). \quad \text{(S3)}$$

is possibly due to out-of-plane tilt of the magnetic field.

We analyze the data at $T$=2 K under high field. It determines $\Delta R_{xy}^{2\pi}(B)$ and coefficient $A$. In this case, we also determine $\Delta R_{xy}^{2\pi}$at 2 K at other magnetic fields as it is proportional to $B$ field. The determined $\Delta R_{xy}^{2\pi}(B)$ and coefficient $A$ at 2 K were also used for other temperatures up to 12 K (the temperature dependence in this range is assumed to be weak), so that the number of the free fitting parameters is reduced.

(Step ii-ii) Then each dataset is fitted to (S1) with free parameters of $\Delta R_{xy}^{\pi}$, $\alpha_{SMR}$, $R_0$, $C$, and $D$ ($\Delta R_{xy}^{2\pi}$ and $A$ are already determined in step (ii-i)).

**Example**: fitting for small signal is shown in Fig.S2D. We fixed to $\Delta R_{xy}^{2\pi}/B$= 0.141 mΩ/T and $A$ = 76.9°. The result is

- $R_0$ = -46.39 mΩ,
- $\Delta R_{xy}^{\pi}$ = 0.49 mΩ,
- $\alpha_{SMR}$ = 65.0°,
- $C$ = -0.010 mΩ/deg,
- $D = -5.47 \times 10^5$ mΩ/deg$^2$.

The $R$ coefficient for the fitting is $R$=0.82.

We note the reason why ordinary-Hall-like signal appears is due to slight out of plane tilt of the B-field while rotating in-plane as mentioned in the main text.

**Supplementary Note 3. Relation between a proposed possible snapshot of local spin configuration relevant for SMR, and the spin quantization or polarization axis $\hat{\boldsymbol{X}}$**

Figure S6 shows a proposed possible snapshot of local spin configuration relevant for SMR. As shown in Fig. 2C, the transverse component of SMR shows minimum at $\alpha = \pi/4$. The quantization axes $\hat{\boldsymbol{X}}$ (see eq.3 and main text) of the spins are mostly perpendicular to the external $B$ field, resulting in $S_x S_y$ (which contributes to SMR (1)) to be always negative.

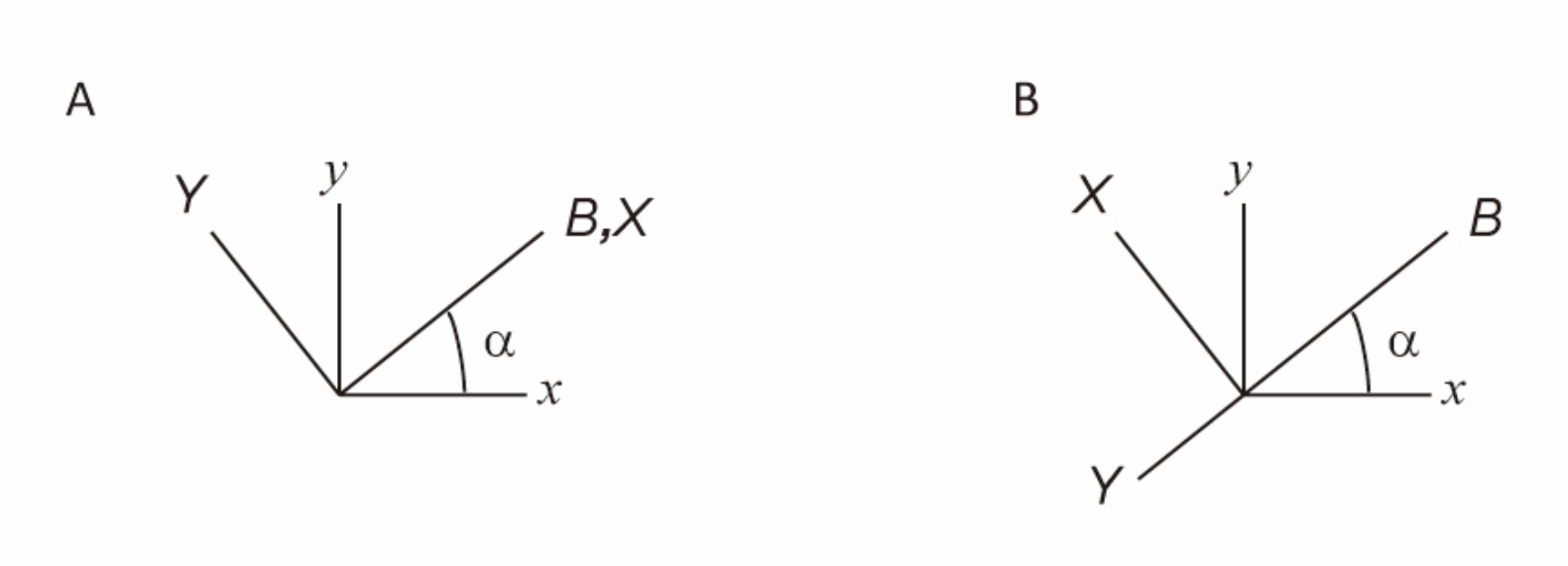

**Fig. S1.**
Notations and directions in Supplementary Note 1: axes $x$, $y$, *X*, *Y*, and magnetic field *B*. Here *X* is the magnetic moment direction. (A), FM case, *X* is along *B*. (B), AFM case. *X* is perpendicular to *B*.

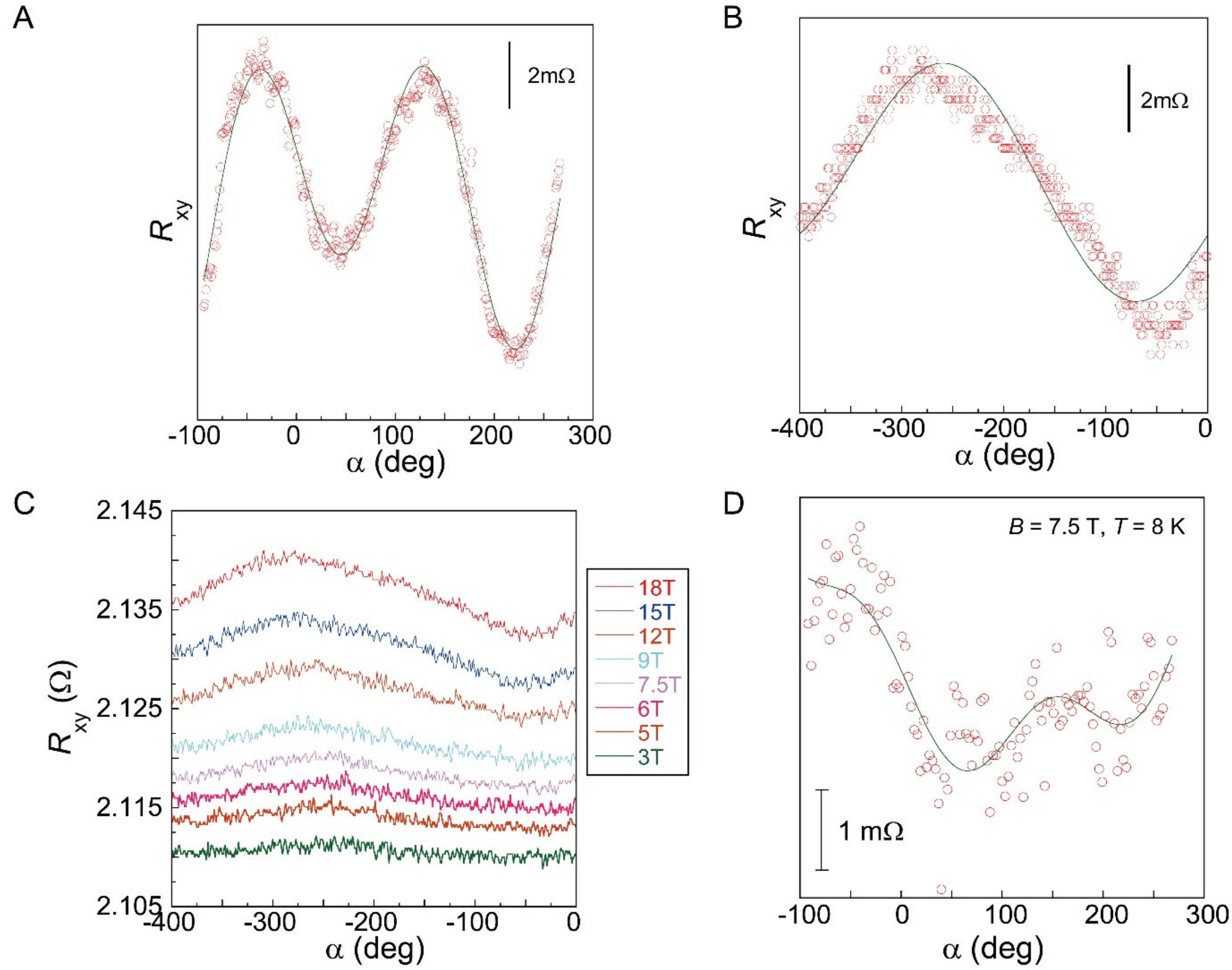

**Fig. S2.**

Angular dependence of $R_{xy}$ for (A), α-$RuCl_3$/Pt at 18T (device #1, same result as in Fig. 2C, the fitting parameters are $R_0$ = 64.5 ± 0.04 mΩ, $\Delta R_{xy}^{\pi}$ = 3.79 ± 0.04 mΩ, $\alpha_{SMR}$ = 45.4 ± 0.3°, $C$ = 0.0069 ± 0.0004 mΩ/deg, and $\Delta R_{xy}^{2\pi}$ = 2.10 ± 0.05 mΩ, assuming $A$ = 30.23°, giving the $R$ coefficient for the fitting $R$=0.98), and for Pt Hall bar at (B), 18 T (a fitting, shown as the solid line, with $\Delta R_{xy}^{2\pi}\cos(\alpha+A)+R_0+C\alpha$ gives $R_0$ = 2.1363 ± 0.000 Ω, $\Delta R_{xy}^{2\pi}$ = 3.2± 0.0 mΩ, $C$ = 0.0069 ± 0.0004 mΩ/deg., and $A$ = 105.4± 0.9°, with the $R$ coefficient for the fitting $R$=0.94) and (C), at various magnetic fields. Data for (A)-(C) were measured at $T$ = 2 K. (D) Fitting of angle-dependent $R_{xy}$ for device #2, at $T$= 8 K and $B$=7.5T. Fitting parameters are $R_0$ = −46.39 mΩ, $\Delta R_{xy}^{\pi}$ = 0.49 mΩ, $\alpha_{SMR}$ = 65.0°, $C$ = −0.010 mΩ/deg, and $D$ = −5.47 × $10^5$ mΩ/$deg^2$, assuming $\Delta R_{xy}^{2\pi}/B$= 0.141 mΩ/T and $A$ = 76.9°. The $R$ coefficient for the fitting is $R$=0.82.

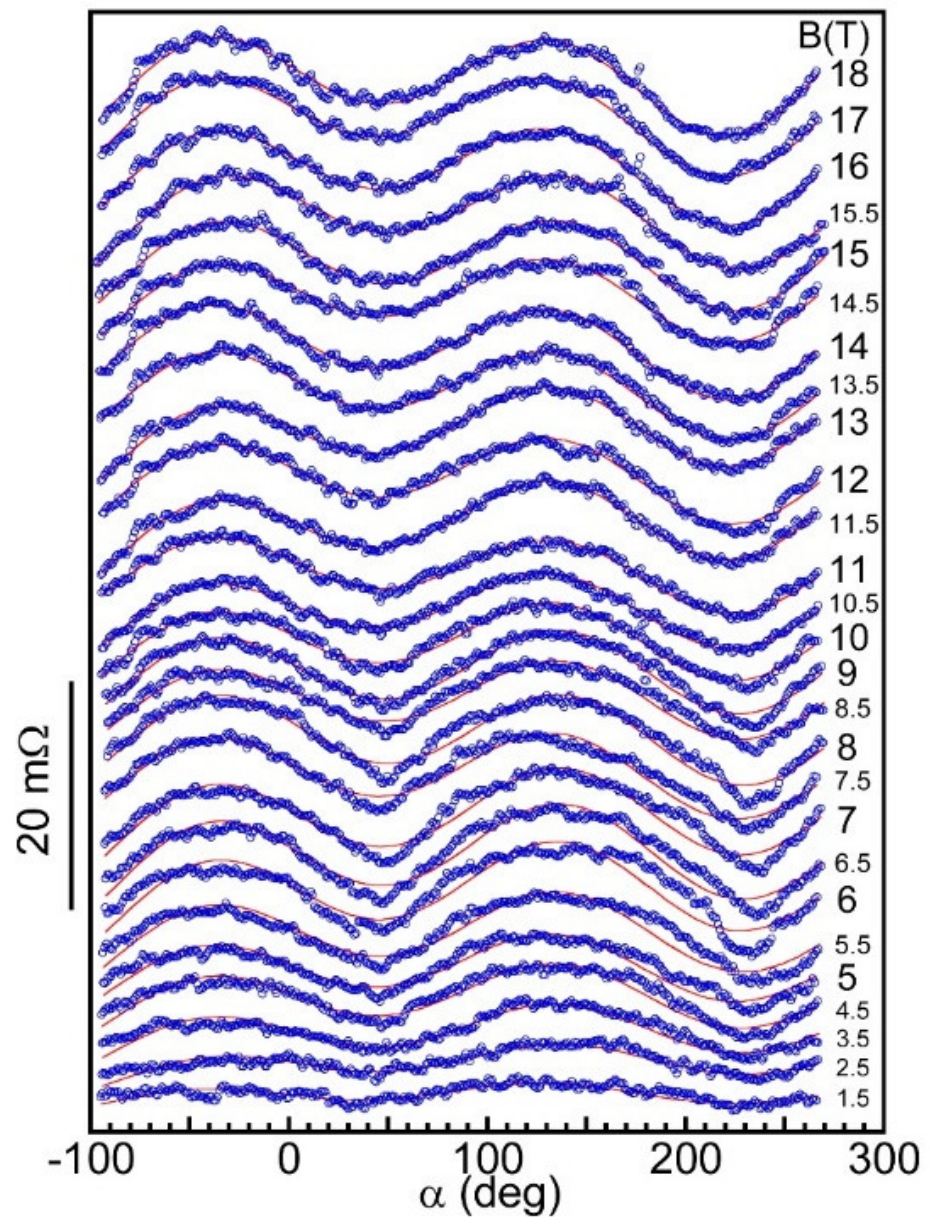

**Fig. S3.**

The magnetic field evolution of the angular oscillation in the Hall resistance ($R_{xy}$) at $T$ = 2 K for device #1. Angular oscillation with a period of $\pi$ is clearly visible. The curves are shifted vertically for clarity. Figure 2B is generated by interpolating these data. The solid red line is obtained by fitting. The scale bar shows 20 m$\Omega$.

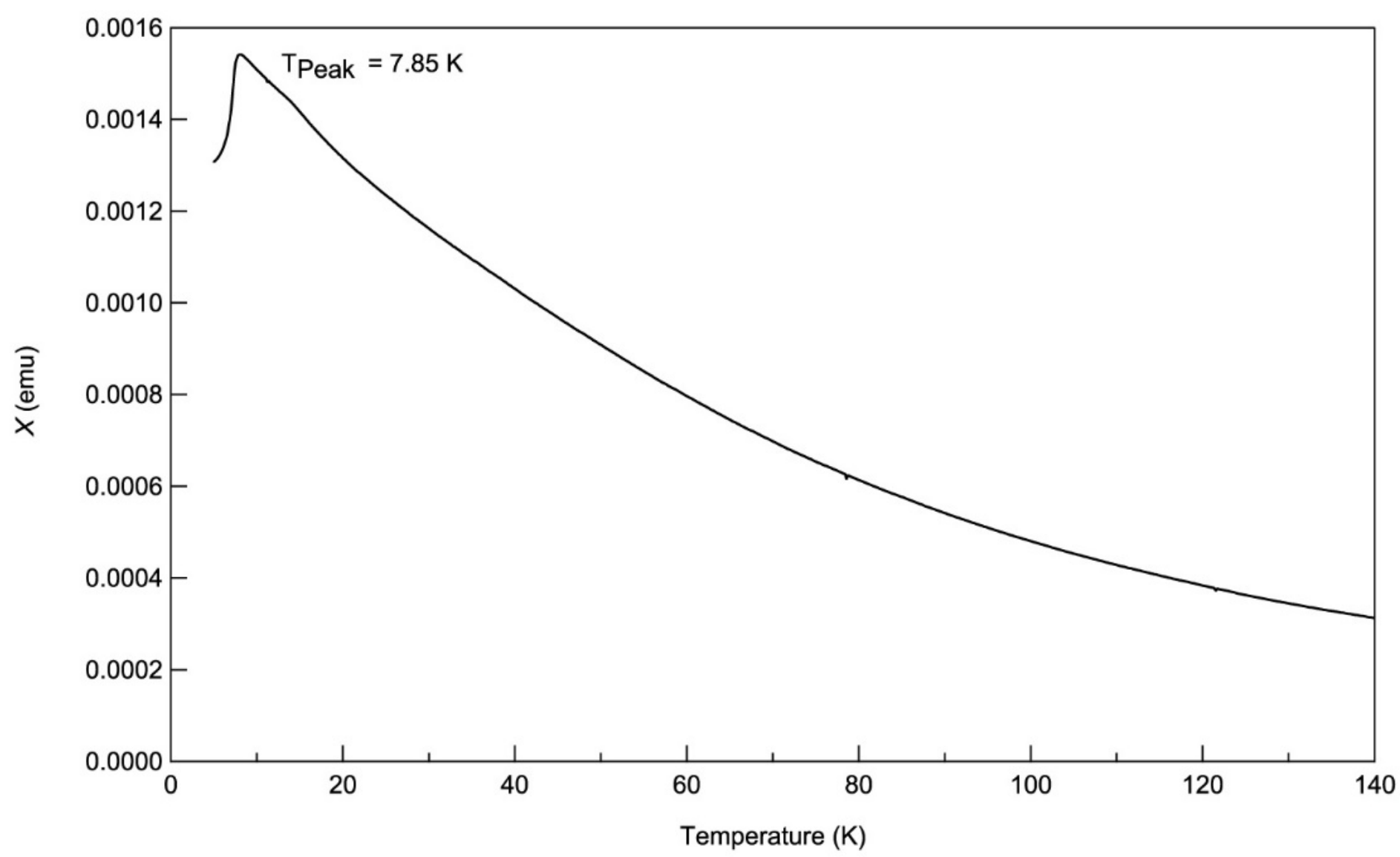

**Fig. S4**

Temperature variation of magnetic susceptibility on $\alpha$-$RuCl_3$ single crystal.

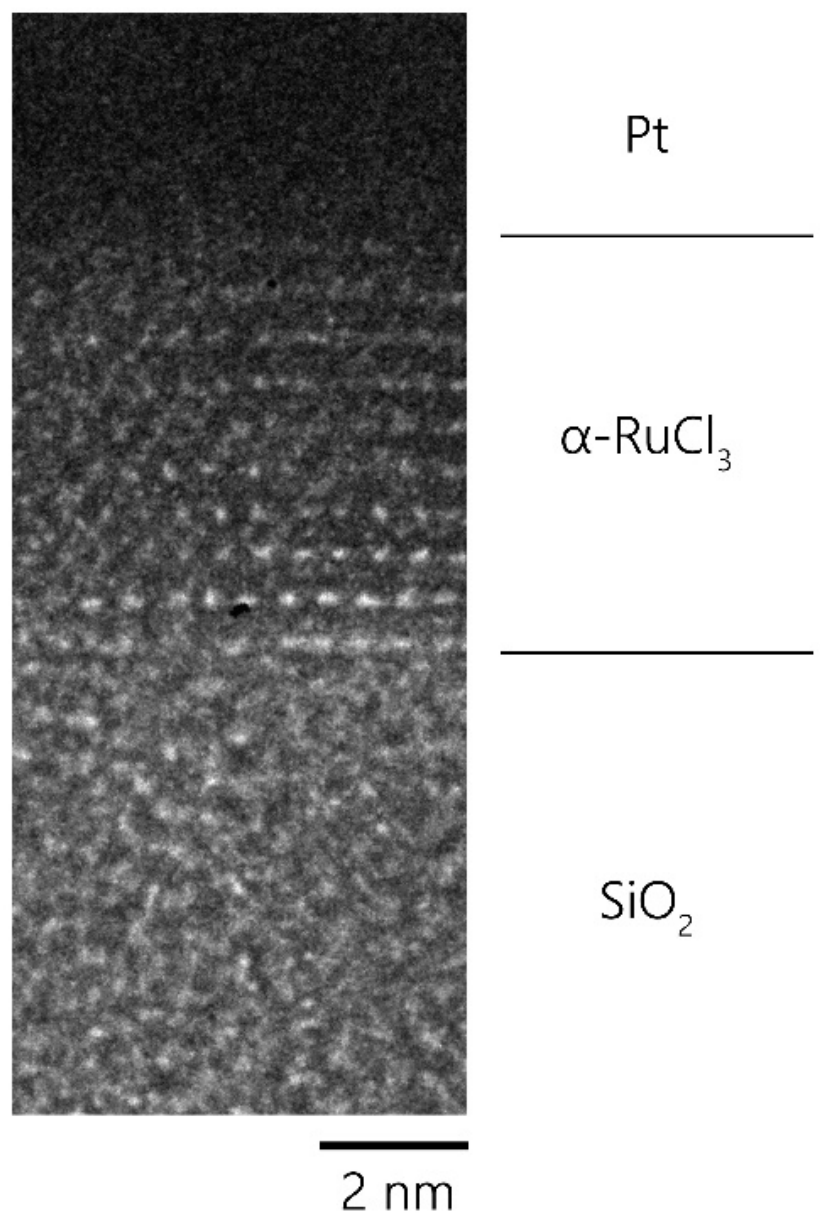

**Fig. S5**

Scanning transmission electron microscopy image for an α-$RuCl_3$/Pt bilayer on Si/$SiO_2$ substrate (Pt and α-$RuCl_3$ thickness is 3 nm and ~ 8 nm, respectively).

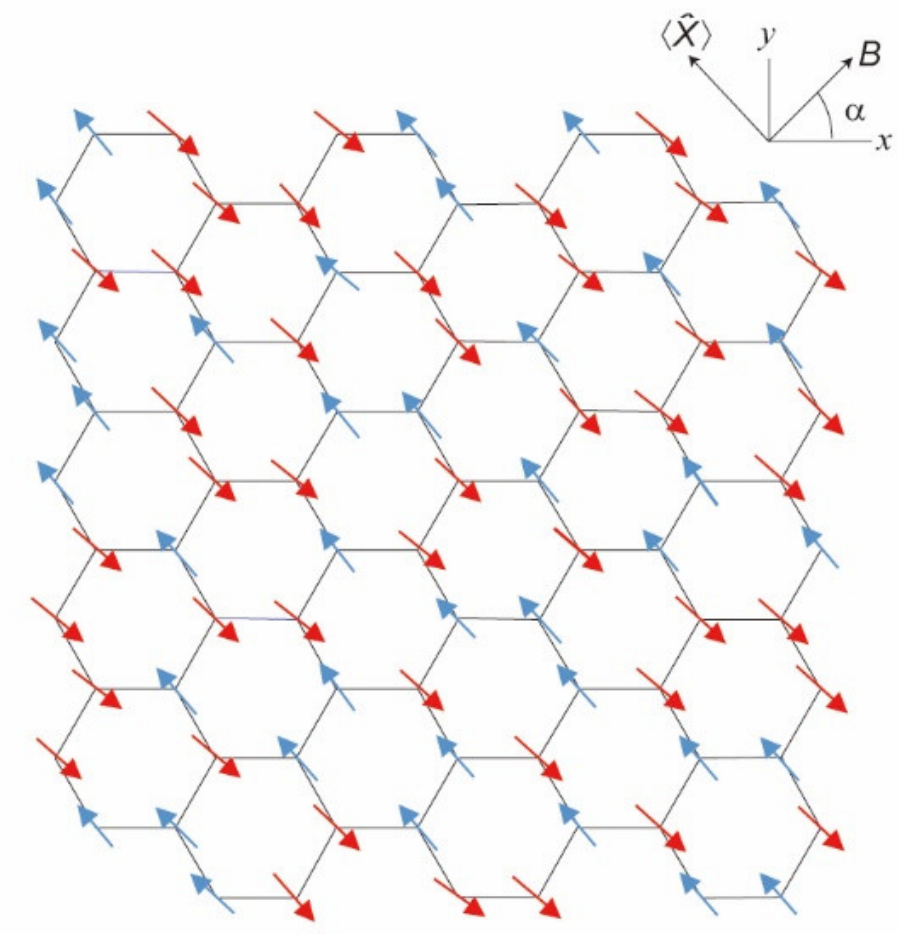

**Fig. S6.**

A proposed possible snapshot of local spin ($S$) configuration relevant for SMR in α-$RuCl_3$ under external in-plane magnetic field $B$. The red and blue arrows indicate spins with mostly opposite directions, but slightly tilted towards the $B$ field direction to give a small longitudinal magnetic moment. $\langle \hat{\boldsymbol{X}} \rangle$ indicates spatial average of $\hat{\boldsymbol{X}}$. In this snapshot, there are equal number of antiferromagnetic and ferromagnetic bonds (two neighboring spins having different or same colors), without any magnetic superlattices. The snapshot also satisfies that $<n> \equiv <S_A - S_B> = 0$ where $S_A$ and $S_B$ are spins on sublattice A and B, i.e. an antiferromagnetic order is not present. Inset of Fig.5 shows another example of similar spin configuration.

**SI References**